# In-memory computing on a photonic platform

Carlos Ríos[1,^], Nathan Youngblood[1,^], Zengguang Cheng[1], Manuel Le Gallo[2], Wolfram H.P. Pernice[3], C David Wright[4], Abu Sebastian[2,*], and Harish Bhaskaran[1,*]

[1]Department of Materials, University of Oxford, Parks Road, Oxford OX1 3PH, UK

[2]IBM Research - Zurich, 8803 Ruschlikon, Switzerland

[3]Institue of Physics, University of Muenster, Heisenbergstr, 11, 48149 Muenster, Germany

[4]Department of Engineering, University of Exeter, Exeter EX4 QF, UK

^These authors contributed equally.

*Corresponding authors. E-mail: ase@zurich.ibm.com, harish.bhaskaran@materials.ox.ac.uk

**Abstract**

**Collocated data processing and storage are the norm in biological systems. Indeed, the von Neumann computing architecture, that physically and temporally separates processing and memory, was born more of pragmatism based on available technology. As our ability to create better hardware improves, new computational paradigms are being explored. Integrated photonic circuits are regarded as an attractive solution for on-chip computing using only light, leveraging the increased speed and bandwidth potential of working in the optical domain, and importantly, removing the need for time and energy sapping electro-optical conversions. Here we show that we can combine the emerging area of integrated optics with collocated data storage and processing to enable all-photonic in-memory computations. By employing non-volatile photonic elements based on the phase-change material, $Ge_2Sb_2Te_5$, we are able to achieve direct scalar multiplication on single devices. Featuring a novel single-shot *Write/Erase* and a drift-free process, such elements can multiply two scalar numbers by mapping their values to the energy of an input pulse and to the transmittance of the device, codified in the crystallographic state of the element. The output pulse, carrying the information of the light-matter interaction, is the result of the computation. Our all-optical approach is novel, easy to fabricate and operate, and sets the stage for development of entirely photonic computers.**

## Introduction

Integrated photonics offers attractive solutions for using light to carry out computational tasks on a chip[1–6] and phase-change materials are emerging as functional materials of choice on photonic platforms[7–15]. On-chip non-volatile memories that can be written, erased, and accessed optically are rapidly bridging a gap towards all-photonic chip-scale information processing[15–17]. However, breaking the dichotomy of the processor and memory as separate units would vastly transform the computing landscape by allowing processing directly on the memory elements—so-called "memcomputing" or in-memory computing. Electronic implementations of such systems, able to carry out complex tasks such as scalar multiplication, bulk-bitwise operations, correlation detection, and compressed sensing recovery, are now emerging[18–22]. It is also suggested that such computational memory machines could solve certain nondeterministic polynomial (NP) problems in polynomial (P) time by exploiting attributes such as inherent parallelism, functional polymorphism and information overhead[23,24]. Photonic implementations of in-memory computing on an integrated photonic chip have the potential to further transform the computing landscape, by providing, ultimately, increased speeds and bandwidths that can come from working directly in the optical domain, leveraging both inherent wavelength division multiplexing capabilities and the rapid technological advances of the Si photonics 'revolution'[25]. Although integrated photonic memories have been showcased in recent years[11–15], carrying out computational tasks in the same device that implements the memory function provides very significant challenges in terms of switching energy, speed, and single shot programming and recovery. In this paper, we overcome such challenges and demonstrate the first instance of a photonic computational memory for direct scalar multiplication of two numbers. Specifically, we demonstrate the multiplication $a \times b = c$, with $a, b, c \in [0,1]$, using a single integrated photonic phase-change memory cell.

Distinct from previous work[17], this multiplication is not achieved by sequential addition, providing a significant advance in computational efficiency. Such a seemingly simple operation lies at the heart of many important computational tasks. For example, an architecture comprising multiple scalar multiplying units could be exploited to calculate, in parallel, the result of the matrix-vector multiplications that underpin key processing operations in the area of 'big data' analytics[18]. For example, such an architecture can be used for the efficient solution of systems of linear equations and a host of other modern-day computational areas, such as machine learning[19] (e.g. for sparse coding approaches to pattern matching of images[19]) and neural networks (e.g. for execution of forward and backward propagation algorithms[26,27]). Our work thus represents a milestone for optical processing in memory.

**Results**

We used the photonic memory device with phase-change materials shown in **Fig. 1a** as the functional element to demonstrate scalar-multiplication of two numbers. To do so, we mapped the numbers to the power of an input pulse $P_{in}$ and the transmittance $T$ of the device, which is set by a *Write* pulse $P_{Write}$, as sketched in **Fig. 1b-d**. Our device relies on the near-field coupling between the propagating electromagnetic mode inside the waveguide and a phase-change material segment (cell) placed on top (of the waveguide) to absorb enough energy to crystallize (anneal over 150°C) or amorphize (melt-quench over ~600°C)[28]. Because most phase-change materials possess a non-negligible imaginary refractive index in the visible and near-IR wavelength range, light is attenuated in different amounts depending on the phase configuration of the material, which gives rise to differentiable transmitted signals, thus encoding information.

The *Write* pulse $P_{Write}$ modulates the transmittance state of the photonic memory device. In particular, we used pulse energies $E_{P_{Write}}$ and $E_{P_{in}}$, such that $E_{P_{Write}} > E_{Th} > E_{P_{in}}$, where $E_{Th}$ is the threshold energy to partially amorphize the phase-change material. As we demonstrate, this simple yet powerful architecture is a promising solution for optical information processing applications, given that light is attenuated without directly blocking the transmission. In our experiments, we utilize the phase-change material alloy, $Ge_2Sb_2Te_5$ (GST) in combination with $Si_3N_4/SiO_2$ photonic waveguides at telecom wavelengths, as sketched in **Fig. 1a** (see methods and materials). To characterize our devices and condition our material for multilevel operation (i.e. establishing specific levels), we used the counter-propagating pump-probe measurement configuration we have previously employed[15]. Pump pulses were used to control the phase-configuration of GST, while a continuous-wave (CW) probe was used to read-out the transmission state. The energy associated with the pump pulse was chosen to be over the energy threshold required to partially amorphize the GST material, whereas the probe was fixed to a lower energy so that transmission could be measured without altering the GST phase-configuration. Multiple and non-volatile levels of transmission were reached, as a result of the mixture between amorphous and crystalline GST[29], by controlling the power of the pump pulse to *Write* (amorphize) up to any level of higher transmission. To *Erase* (recrystallize) we used two approaches: a train of decreasing-energy pulses as employed in[10,15] and, as explained later in this paper, a single two-step pulse. The former allowed us to reach any of the lower transmission levels, while the latter allowed us to recrystallize down to the baseline. Whilst using the train of pulses, there are no restrictions in intra-level transitions, i.e., any level can be reached starting from any other and the number of levels is limited by the geometry of the GST cell and the available signal-to-noise ratio.

**Fig. 1b** shows the pulse and transmittance function that represent the two numbers which require multiplication. A *Write* pulse, $P_{Write}$, was used to program a specific level of transmission $T$ of the device, as illustrated in **Fig. 1c**, which relies on the multilevel conditioning of the material. The portion of the $P_{Write}$ pulse that is transmitted was not taken into account (in the calculation of the result of the multiplication). On the other hand, the read pulse $P_{in}$, representing the second number, propagates through the photonic memory device, experiencing a transmittance given by the current phase-configuration of the GST cell (as conditioned by pulse $P_{Write}$) but does not induce any change to the material, as shown in **Fig. 1d**. The output pulse of $P_{in}$, $P_{out} = T(P_{Write}) \times P_{in}$, is the result of the multiplication by mapping the multiplicand $a$ to $T$ and the multiplier $b$ to $P_{in}$. Such a direct multiplication process is considerably more efficient than previous demonstrations of multiplication by sequential addition[17,30–32].

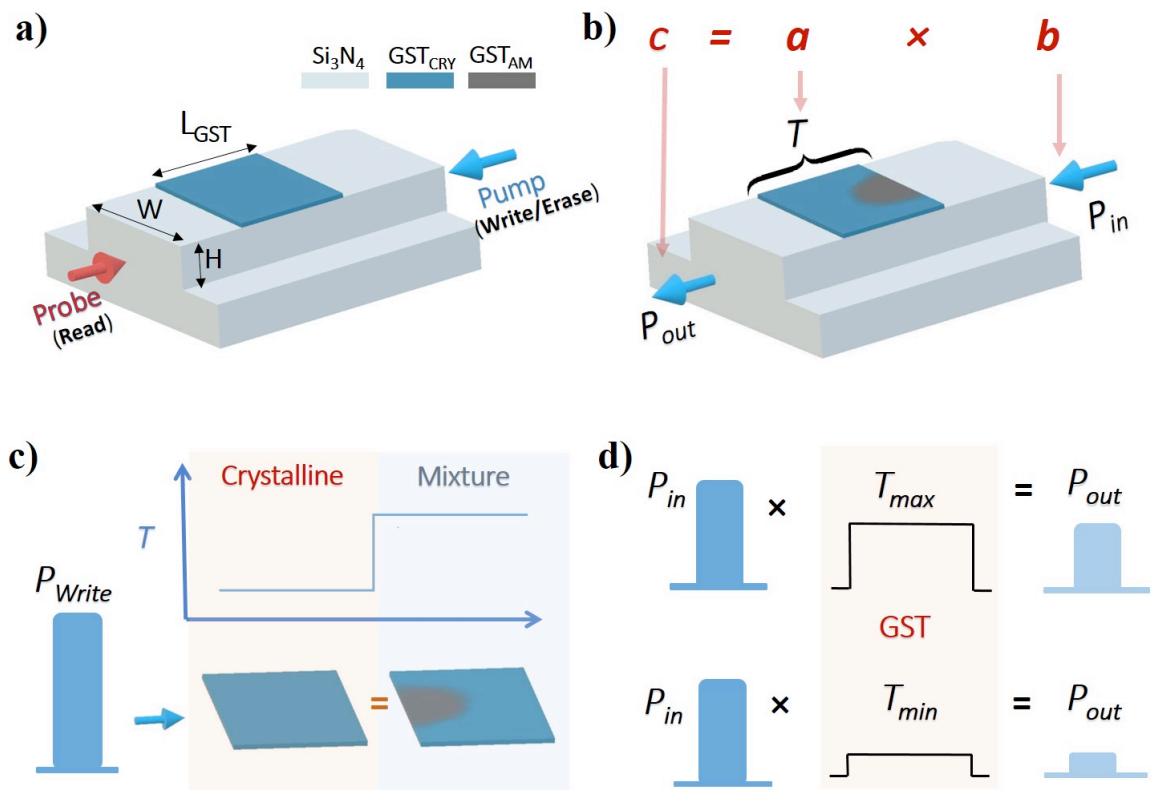

**Figure 1. GST-based photonic memory cell and scalar multiplication concept** a) Scheme of an on-chip bidirectional pump-probe to both switch and readout the transmission in a GST-based nanophotonic memory device. Pump consisted of pulses at $\lambda = 1590\ nm$ and the probe, of a CW at $\lambda = 1598\ nm$. $L_{GST} = 1\ or\ 2\ \mu m$, $W = 1.3\ \mu m$, and $H = 165\ nm$ (etched down from a 330 nm thick $Si_3N_4$). b) Scheme of the multiplication of two scalars $a$ and $b$, codified in the material transmittance $T$ and in the energy of the input pulse $P_{in}$, as detailed in c) and d). c) The first pulse $P_{Write}$ is used to switch the GST between levels, thus defining the transmittance $T$ of the device. d) A less energetic input pulse, $P_{in}$ which propagates through without inducing phase-change but being modulated by $T$, set by $P_{Write}$ in the previous step.

We first optimized the parameters of our devices, particularly to decrease the energy consumption and to improve the speed of operation. For a chip with the 1 µm-long cell of GST, the pump pulse width was varied while keeping a constant power to reach different transmission levels, as shown in **Fig. 2a**. For pulses with widths longer than 45 ns there is a saturation due to the finite size of our memory cell, in which no more amorphous material can be obtained without using higher powers. Therefore, longer pulses represent a waste in energy as the GST will not further amorphize and could even be ablated. From these results, it can also be observed that 25 ns pulses induce a change equivalent to 75% of the maximum achievable transmission. Indeed, we found that for cells between 1 and 4 µm long, 25 ns pulses were enough to achieve clear transmission contrast among levels, limited only by the SNR of the measurement. We further reduced the energy consumption by using single shot *Erase* pulses from any arbitrary level to the fully crystalline level (baseline). This was achieved by using the double-step pulse sequence shown in **Fig. 2b**. This pulse, with a total switching energy of approximately 577 pJ, is sufficiently energetic and long to induce the requisite recrystallization (and represents an improvement by a factor of more than 100 in terms of energy and 25 in terms of speed cf. previous work[15]). In terms of operational speed for a *Write/Erase* cycle, it can be observed in **Fig. 2b** that approximately 200 ns (including the pulse duration) is required to obtain a stable transmission level for both the *Write* and the *Erase* pulse. However, this time varies with the pulse width: the shorter the pulse, the shorter it takes for the material to cool down and for the refractive index to stabilize, the timescale of which is governed by the thermo-optical effect[33]. With a pulse separation of 200 ns and taking into account the cooling time for the *Erase* pulse, the device can truly be operated at 2.5 MHz, even though the effect of the time separation between *Write* and *Erase* pulses on the level transitions have not been optimized. Further improvements in terms of speed and energy can be achieved by decreasing the pulse width to even shorter times, especially the *Write* pulse which could be a pico- and even femto-second pulse[10,34,35], thus reducing the dead time between the two pulses (which are used to perform the scalar multiplication—see later).

Since the ability to program our integrated phase-change photonic cells into multilevel states is key to the implementation of direct scalar multiplication, we demonstrate in **Fig. 2c** and **2d**, multilevel behavior in a 2 µm long GST cell using single-shot *Write* and *Erase* with the same pulse parameters of **Fig. 2b**. Thirteen transmission levels were programmed, limited only by the noise of the photodetector. We found a linear response of the attenuation caused by the GST cell as the switching energy is increased. For large energies, the linearity is lost due to saturation, which implies that the GST is in its most amorphous state. The beginning of such saturation can be observed in **Fig. 2c** for switching energies larger than 210 pJ. For this reason, only energies up to 220 pJ were considered. To study the error in reaching the programmed transmission level, we sent combinations of *Write* and *Erase* pulses to reach ten different transmission levels in twenty consecutive sequences. This experiment was repeated three times. The results of subtracting the actual transmission level after a particular transition from the originally programmed level are plotted in the **Fig. 2e**. The programmed levels were taken as the average level after several back-and-forth switching during a conditioning process[15]. With a standard deviation of 0.35 % (for the total change in transmission for $\Delta T$) a confidence interval for each transmission level can be established so that each level is uniquely distinguished. This value is determined by SNR but also by the variation in power of the pump pulses due to fluctuations in the electro-optical conversion.

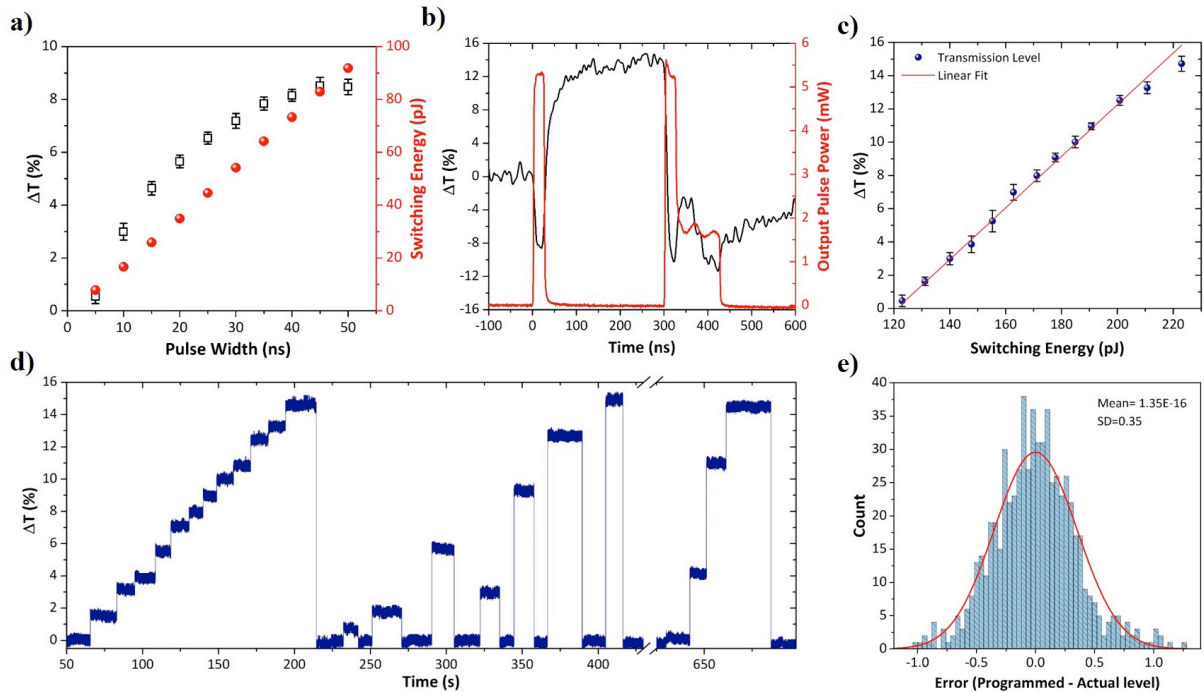

**Figure 2. Device optimization.** a) Pump pulses with varying widths achieve different transmission levels before saturating at around 45 ns, and their equivalent switching energy (energy absorbed by GST). Shorter pulses can be chosen, thus saving energy and time during memory operations. $\Delta T = (T - T_{min})/T_{min}$ is the change in transmission of the level *T* with respect to the baseline $T_{min}$. b) Single-shot *Write* and *Erase* pulses. The output pulse power corresponds to the power after propagating through the GST cell. An amorphization to a higher transmission level is achieved with a single square pulse[15], while a new *Erase* double-step pulse is used to fully recrystallize. The latter consists of 14.1 mW (or peak power) for 25 ns and 5.64 mW (0.4 times the peak power) for 100 ns, where peak power is the power used to reach the maximum transmission level. c) Switching energies used for the multilevel conditioning shown in d) with a linear response by the GST memory cell. d) Multilevel conditioning of a 2 µm-long GST cell. Thirteen clearly distinguishable levels are demonstrated and accessed randomly, the number of levels is limited only by the SNR and the confidence interval as shown in e). e) Error of the multilevel operation calculated from subtracting the measured transmission level from the programmed level in 600 transitions or switching. The red line corresponds to a normal distribution fitting curve.

For the proper execution of scalar multiplication operations using multilevel memory states, it will also important for such states to be stable over time. Thus, we show in **Fig. 3** the optical transmission traces for up to $10^4$ s, measured with the aim of studying the level stability and thus, the optical drift. Transmission levels in an arbitrary order were written and erased in two 2 μm-long GST memory cells. Subsequently, the devices were kept in an intermediate transmission level for a prolonged period of time, one having a 0.1 mW probe ON and the other OFF, as shown in **Fig. 3a** and **Fig. 3b**, respectively. Mechanical drift of the sample stage was observed as result of the relaxation of the picomotors over time; however, once the chip was placed in a stable position, the transmission level remained constant for the case in which the probe was kept ON. Therefore, we do not find measurable drift for up to $10^4$ s of constant measurement, a property that we attribute to the fact that the crystalline phase, which is more stable than the amorphous, determines and dominates the optical absorption. This drift-free process represents a big advantage for photonic computational memory over its electrical analog which undergoes very significant resistance drift over time, thus preventing (or at least making it very difficult) to achieve and maintain reliable levels[36–39]. Moreover, the same specific transmission levels were retrieved after the measurement, hence the multilevel conditioning is also preserved over time. On the other hand, for the case when the probe was turned OFF, a drift of nearly 9% was observed once the probe was turned on again. This drift is due to the relaxation of the material when the probe is removed, given that the probe itself heats up the material to a constant temperature to cause a thermo-optical effect that modifies the values of the complex refractive index without crystallizing[33]. However, once the probe was turned ON, the drift is easily corrected for by sending a write pulse with the same energy as that required to reach the originally programmed level (i.e the level set immediately prior to the probe being turned off). As result, we are able to reliably retrieve both the original level and any other level, demonstrating that the multilevel conditioning was preserved. We were also able to avoid the drift entirely by simply decreasing the probe power to 0.05 mW—half of its previous value. In this case, the material relaxation after the probe is turned OFF and then ON is negligible, given that the temperature is lower at the GST, as shown in **Fig. 3c**.

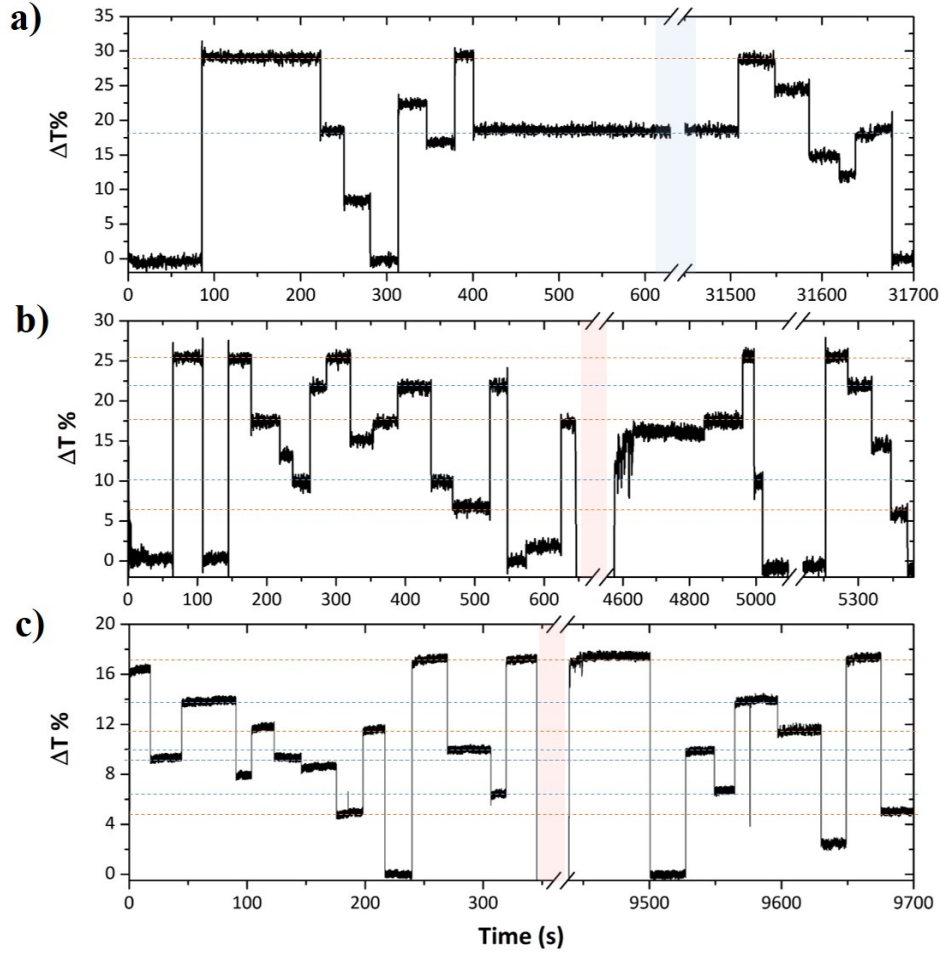

**Figure 3. Drift in transmission.** a) Multilevel operation of a 2 µm-long GST memory cell using 50ns pulses with energies in the range of 350-600 pJ to *Write* (upward transitions) and a train of power-decreasing pulses to *Erase* (downward transitions) level[15]. The blue highlighted area corresponds to ~8.5 h of constant measurement with a CW probe of 0.1 mW (inside the waveguide). b) Same as a) but turning the probe OFF for around 1.5 h. A drift is observed between 4600 and 4800 s, which is corrected by sending the pulse energy of the level where the memory was originally set. c) Multilevel operation of a different 2 µm-long GST memory cell using 25 ns pulses with energies in the range of 200-360 pJ (see **Fig. 2d**). The *Write* and *Erase* were done in the same way as for a) and b). The probe was turned off for a time of ~2 h but in this case, the CW probe power was reduced to 0.05 mW, which is enough to avoid drift. The colored traces are added to track the evolution of the levels.

We now demonstrate the multiplication $a \times b = c$, with $a, b, c \in [0,1]$. We used the same photonic memory device with 2 µm-long GST cell and the single-shot *Write* and *Erase* scheme. We mapped the multiplicand $a$ to the transmittance of the device, $T = \Delta T + T_0$, where $\Delta T$ corresponds to the linear response of the change in transmission as function of the $P_{Write}$ pulse (shown in **Fig. 2d**) and $T_0$ is the baseline transmission level (fully crystalline). Subsequently, the multiplier $b$ is mapped to the energy of a second pulse $P_{in}$. The result of the multiplication is calculated from the output of this latter pulse, which is equivalent to $P_{out} = T \times P_{in}$. We generated both $P_{Write}$ and $P_{in}$ in the same manner by tuning the pulse power at the

EOM (see materials and methods). **Fig. 4a** shows the input pulse energies used for both multiplicand and multiplier, highlighting the energies that will induce phase switching. The measured energy corresponds to the output pulse energy after propagating past the GST cell and through one grating coupler. In particular, we use $P_{Write} \in [180\ pJ, 354\ pJ]$ or equivalently, $T \in [0, 0.143]$ (see **Fig. 2d**), and $P_{in} \in [0\ pJ, 112.8\ pJ]$. The linear response of the memory device demonstrated in **Fig. 4a** is highly convenient as it allows for easy mapping of a scalar to the pulse energy without relying on fitting functions[18]. In **Fig. 4b**, we show the experimental realization of three multiplications and the actual output pulses, $P_{out}$. In particular, we used the minimum and the maximum of the multiplicand: $T_{min} = 0$ (i.e. $a = 0$) and $T_{max} = 0.143$ (i.e. $a = 1$), and the maximum and an intermediate value for the multiplier: $P_{in}^1 = 112.8$ pJ (i.e. $b = 1$) and $P_{in}^2 = 38.9$ pJ ($b = 0.4$). In this figure, it can be observed that the output powers have distinct energies, which can be measured and then rescaled to obtain the result of the operations: 1×0 = 0, 1×1 = 1, and 1×0.4 = 0.4. Note that the lowest transmission level of the three, which in theory corresponds to 0.4, is actually smaller than the level for 0, using $P_{in}^1$. This is due to the fact that even in the fully crystalline state, GST will not absorb all the light, unless a longer GST cell (than here) is used to increase the optical attenuation[14]. Therefore, there is an offset given by the transmission baseline $T_0$, which has to be subtracted from every multiplication to enable exact linear rescaling to the $[0, 1]$ results. In **Fig. 4c**, we demonstrate 429 multiplications choosing arbitrary values for *a* and *b* with the associated error (difference between the exact and the measured value of *c*) shown in **Fig. 4d**. We found good agreement between the exact and the measured value of the multiplication, having an error that spreads as *a* and *b* get larger, as shown in the inset of **Fig. 4d**. The exact value was calculated from the linear fits in the characterization of the device and the subsequent mapping to $[0, 1]$ for both multiplicand and multiplier. The measured value corresponds to the average of the output pulse energy, correcting the offset, and normalizing to $[0, 1]$. While the results of the multiplication are not exact due to factors such as the fluctuations in the values of *T*, as shown above in **Fig. 2e**, this kind of multiplication operation has proved useful in application areas such as machine learning[19]. Moreover, in application domains where arbitrarily high accuracy is required, ideas such as mixed-precision computing can be employed where the low precision multiply unit is used in conjunction with a high precision unit[18]. Moreover, the non-volatility of the levels implies that the multiplicand is fixed until the next pulse excitation changes its state. This property, in an architecture comprising several scalar multiplying units, could be exploited for efficient calculation of matrix-vector multiplications or solutions to systems of linear equations using iterative Krylov subspace methods, such as the Conjugate Gradient or the Generalized Minimum Residual[40]. This type of matrix-vector multiplication is well suited to our photonic framework and a proposed architecture to implement this. Using such methods, an input matrix A could be mapped to GST nanophotonic memory cells and the solution $x$ of $Ax = b$ with $A \in \mathbb{R}^{N \times N}$ and $b, x \in \mathbb{R}^N$ can be computed from iterative algorithms, as demonstrated for electrically switchable phase-change memory devices[18].

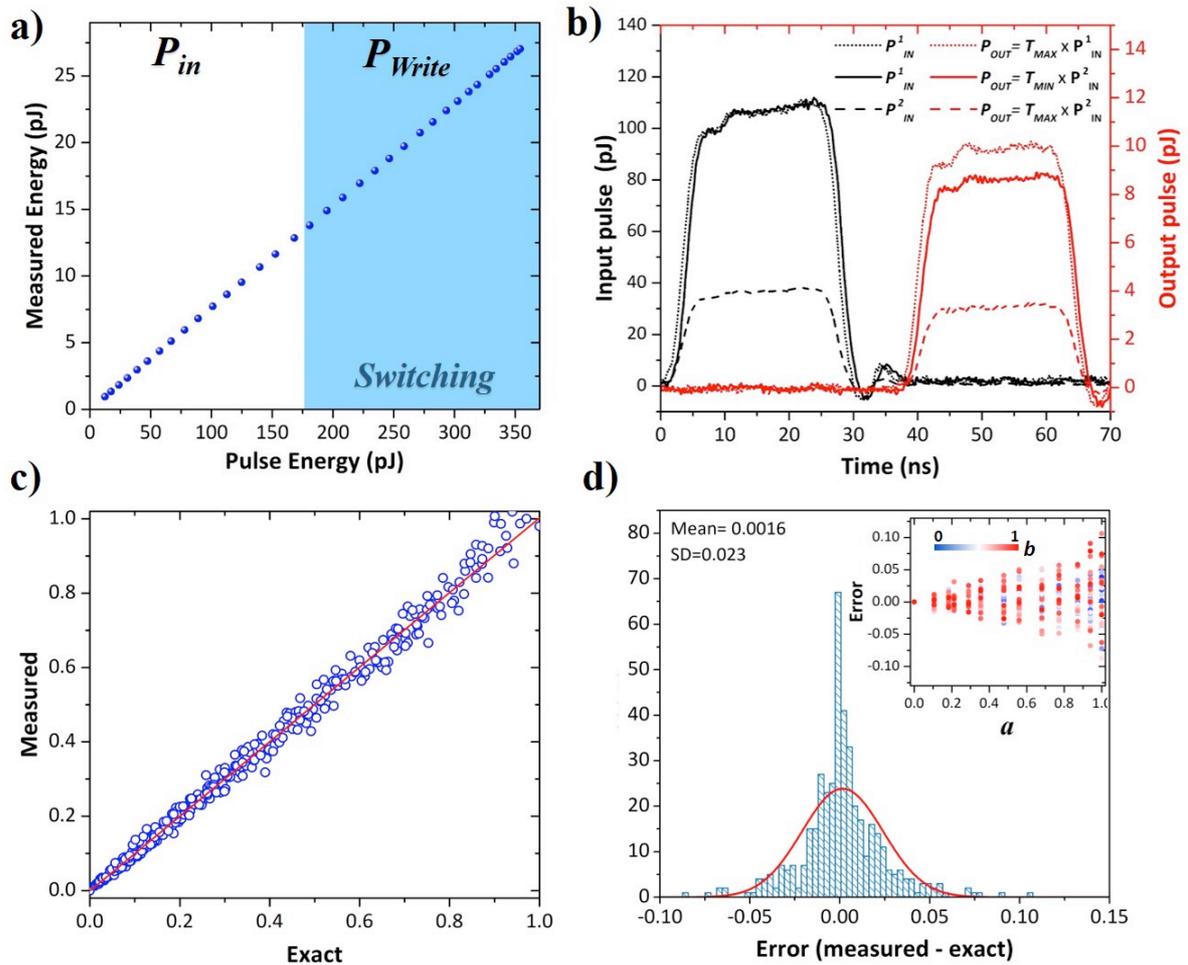

**Figure 4. Scalar multiplication.** a) Input pulse energies for $P_{Write}$ and $P_{in}$ and corresponding measured energy (past the GST cell and a grating coupler). All the measurements were carried out with the memory cell in the transmission baseline of the multilevel conditioning shown in **Fig. 3c**. The highlighted area corresponds to the pulse energies that induce switching to a partial amorphous state (after the switching the memory was *Erase* down to the baseline again before the next pulse). b) Demonstration of three multiplications using two different $P_{in}$, $P^1_{in} = 112.8$ pJ and $P^2_{in} = 38.9$ pJ, to reach the maximum ($T = 0.143$) and/or the minimum ($T = 0$) of the transmittance. The result when multiplying by $T = 0$ corresponds to the level-specific offset. The time delay between pulses is due to the difference in optical path between the reference pulse and the pulse that is coupled into the photonic chip, both obtained from the pump pulse using a 90/10 beam splitter. c) realization of 429, $c = a \times b$ multiplications equivalent to 13 different, equally spaced, values for $T$ (reached with $P_{Write} \in [180\ pJ, 354\ pJ]$) thus, creating 13 values for the multiplicand $a$, and 33 different values for $P_{in} \in [0\ pJ, 112.8\ pJ]$ for 33 scalars corresponding to $b$ (*Note: $a, b, c \in [0, 1]$*). $P_{Write}$ was sent first to establish a level followed by the 33 $P_{in}$ pulses, before changing levels again. d) Error of the scalar multiplication calculated from subtracting the measured value from the exact. The red line corresponds to a normal distribution fit whose Mean and Standard Deviation (SD) are shown on the plot. The inset corresponds to the error as function of the values of *a* and *b,* this shows how for values closer to 1, the error grows linearly, which in turn explains the spread for values close to 1 in c).

**Discussion**

We have demonstrated direct multiplication of scalar numbers using photonic in-memory computing. This is achieved by using the distinct interaction of two pulses, each of which represents a number to be multiplied—one with energy above the switching threshold of a phase change material, and another one below—in an integrated photonic waveguide. We show a linear response in our photonic device, which is an important feature of our approach and avoids using fitting functions, as in the case of pseudo-ohmic electrical phase-change memory devices[18]. Furthermore, we have introduced a new *Erase* mechanism in which only a single pulse is required, which when compared to the train of energy decreasing pulses used in Ríos *et al.*[15] and Stegmaier *et al.*[10], is an important advance, as the *Erase* time is reduced by orders of magnitude (from microseconds to 125 ns) and the *Erase* energy reduced from tens of nJ to the sub-nJ region. When implementing our device in a matrix-vector multiplication architecture, the entries for the vector can be fixed values codified into the phase change material with sub-nJ energy-consumption. However, this process is realized only once, given the non-volatile nature of such materials. The entries for each row of the matrix, on the other hand, are codified into the energy of input pulses. Thus, any pulse below the threshold switching can be used if the output pulse is resolved by a detector, leading to low energies of even fJ. The energies of the input pulses determine the energy consumption for each scalar multiplication once the PCMs have been written, a method that could significantly reduce the memory access power. While in terms of energy consumption our approach is expected to perform close to its electronic analogous[18], we have demonstrated that the photonic in-memory computing is a drift-free and linear process, which represent very significant advantages even beyond those provided by using light as information carrier. The most significant advantage though is the ability to process optically transmitted signals without having to employ highly inefficient electro-optical converters.

Moreover, the change in transmission saturates when increasing the width of the pump pulse (with fixed power); we find that we can optimize the pulse length such that a single 25 ns pulse is sufficient to reach distinguishable levels from the fully-crystalline (baseline) state in 1 and 2 μm-long GST cells. Finally, we have studied the reproducibility of the multiple levels, the optical drift, and the noise in our system to conclude that the photonic GST cell has no relevant effect in transmission fluctuations for durations up to $10^4$ s. The SNR, the errors in achieving the programmed level, and the noise are ultimately limited by the quality of the electronics used in the system, and not inherent to our device. These results confirm the potential of phase-change materials in photonic hardware computational paradigms—including the ability to perform computations with memory in the same physical location, using light. While there is plenty of room for improvement, these results are at the vanguard of collocated memory and processing on a photonic platform.

**Methods**

*Sample Fabrication:* The nanophotonic memory cells were fabricated on 330 nm $Si_3N_4$/3.3 µm $SiO_2$ wafers. A JEOL JBX-5500ZD 50kV Electron-beam lithography (EBL) was used to write the photonic circuitry using MaN-2403 negative resist, followed by a reflow process of 90s at 100°C. Subsequently, reactive ion etching (RIE) in $CHF_3/Ar/O_2$ was carried out to etch 165 nm of the $Si_3N_4$ and thus obtain the bare photonic device. A second EBL writing step using poly(methyl methacrylate) (PMMA), followed by a lift-off process, was used to pattern the phase-change materials. A stack of 10nm of GST with a 10nm ITO capping (to avoid oxidation) was deposited in an argon environment using a homebuilt RF sputtering system (Nordiko). Before the measurements the GST was crystallized on a hotplate following a 5 minute anneal at 250°C.

*Measurement Setup:* A pump-probe experimental setup was used to carry out the multilevel, time-resolved, and multiplication measurements. The optical signals were confined to the photonic circuit; that is, *Write*, *Erase*, read-out, and multiplication were all realized within the integrated chip. To avoid interference, two different C+L CW tunable lasers sources were employed, wavelengths of 1598 nm (TSL-550, Santec) and 1590 nm (N7711A, Keysight) were chosen for the probe and pump, respectively. The pump pulses - as well as the multiplicand pulse $P_{in}$ - were subsequently generated with an electro-optical modulation (Lucent Technologies, 2623NA), which was controlled by a 100Mhz electrical pulse generator (AFG 3102C, Tektronix). The pulse was further power amplified by a low-noise erbium-doped fiber amplifier (AEDFA-CL-23, Amonics). Both the pump pulses and the probe were coupled into the photonic device using integrated gratings couplers with transmission peak at 1598 nm and coupling efficiencies of ~20%. The counter-propagating scheme was used to ease the separation of the signals and tunable optical filters (OTF-320, Santec) were introduced to the optical lines to further suppress noise resulting from reflections. At one output of the device, the CW probe was divided into two beams using a 90/10 beam splitter to measure the time-resolved and the long-term transmission with a 200KHz low-noise photoreceiver (NewFocus, 2011) and a 125 MHz photodetector (New Focus, 1811), respectively. At the other output, the transmitted pulses were monitored using a 1 Ghz photodetector (NewFocus, 1611).

**Acknowledgements**

This research was supported by EPSRC via grants EP/J018694/1, EP/M015173/1 and EP/M015130/1 in the UK; the Deutsche Forschungsgemeinschaft (DFG) grant PE 1832/2-1 in Germany; and the European Research Council grant 682675. HB thanks helpful discussions with A. Ne. All authors thank the collaborative nature of European science for allowing this work to be carried out.